\documentclass[twocolumn,showpacs,preprintnumbers,amsmath,amssymb,superscriptaddress,prb]{revtex4}
\usepackage{graphicx} 

\begin{document}

\date{\today}

\preprint{APS/123-QED}

\title{Magneto-optical spectroscopy of (Ga,Mn)N epilayers}

\author{S.~Marcet}
\affiliation{"Nanophysique et Semiconducteurs",
CEA-CNRS-Universit\'{e} Joseph Fourier Grenoble, Laboratoire de
Spectrom\'{e}trie Physique, BP~87, 38402 St Martin d'H\`{e}res
cedex, France.}

\author{D.~Ferrand}\email{david.ferrand@ujf-grenoble.fr}
\affiliation{"Nanophysique et Semiconducteurs",
CEA-CNRS-Universit\'{e} Joseph Fourier Grenoble, Laboratoire de
Spectrom\'{e}trie Physique, BP~87, 38402 St Martin d'H\`{e}res
cedex, France.}

\author{D.~Halley}\altaffiliation[Permanent address: ]{IPCMS-GEMME, 23 rue du Loess, BP~43, 67034 Strasbourg cedex 2, France.}\affiliation{"Nanophysique et Semiconducteurs",
CEA-CNRS-Universit\'{e} Joseph Fourier Grenoble, Laboratoire de Spectrom\'{e}trie Physique, BP~87, 38402 St Martin d'H\`{e}res cedex, France.}

\author{S.~Kuroda}\altaffiliation[Permanent address: ]{Institute of Materials Science, University of Tsukuba, Tsukuba, Ibaraki 305-8573, Japan.}\affiliation{"Nanophysique et Semiconducteurs",
CEA-CNRS-Universit\'{e} Joseph Fourier Grenoble, Laboratoire de Spectrom\'{e}trie Physique, BP~87, 38402 St Martin d'H\`{e}res cedex, France.}

\author{H.~Mariette}
\affiliation{"Nanophysique et Semiconducteurs", CEA-CNRS-Universit\'{e} Joseph Fourier Grenoble, Laboratoire de Spectrom\'{e}trie Physique, BP~87,
38402 St Martin d'H\`{e}res cedex, France.}

\author{E.~Gheeraert}
\affiliation{Laboratoire d'Etudes des Propri\'{e}t\'{e}s \'{e}lectroniques des Solides, CNRS, BP 166, 38042 Grenoble cedex 9, France.}

\author{F.J Teran}\altaffiliation[Permanent address: ]{Dpto. F\'isica de Materiales, Universidad Aut\'onoma de Madrid, 28049 Madrid, Spain}\affiliation{Laboratoire de Champs Magn\'{e}tiques Intenses, CNRS, BP 166, 38042 Grenoble cedex 9, France.}

\author{M.L~Sadowski}\affiliation{Laboratoire de Champs Magn\'{e}tiques Intenses, CNRS, BP 166, 38042 Grenoble cedex 9, France.}

\author{R.~M.~Galera}\affiliation{Laboratoire Louis
N\'{e}el, CNRS, BP 166, 38042 Grenoble cedex 9, France.}

\author{J.~Cibert}\affiliation{Laboratoire Louis
N\'{e}el, CNRS, BP 166, 38042 Grenoble cedex 9, France.}

\begin{abstract}
We report on the magneto-optical spectroscopy and
cathodoluminescence of a set of wurtzite (Ga,Mn)N epilayers with a
low Mn content, grown by molecular beam epitaxy. The sharpness of
the absorption lines associated to the Mn$^{3+}$ internal
transitions allows a precise study of its Zeeman effect in both
Faraday and Voigt configurations. We obtain a good agreement if we
assume a dynamical Jahn-Teller effect in the 3d$^{4}$ configuration
of  Mn, and we determine the parameters of the effective
Hamiltonians describing the $^{5}T_{2}$ and $^{5}E$ levels, and
those of the spin Hamiltonian in the ground spin multiplet, from
which the magnetization of the isolated ion can be calculated. On
layers grown on transparent substrates, transmission close to the
band gap, and the associated magnetic circular dichroism, reveal the
presence of the giant Zeeman effect resulting from exchange
interactions between the Mn$^{3+}$ ions and the carriers. The
spin-hole interaction is found to be ferromagnetic.
\end{abstract}

\pacs{75.50.Pp, 75.30.Hx, 78.20.Ls}

\keywords{Suggested keywords}

\maketitle

\section{Introduction}\label{sec:level1}
Extrapolating the Zener model of carrier-induced ferromagnetism to
wide bandgap diluted magnetic semiconductors (DMS) predicts that
high critical temperatures should be achieved, provided several
demanding assumptions are satisfied. In particular, $p$-type
(Ga,Mn)N would be ferromagnetic with a critical temperature
exceeding room temperature. This implies however the incorporation
of 5\% of Mn into GaN, substituting Ga in the form of
$\text{Mn}^{2+}$ ions, and a strong $p$-type doping.\cite{Dietl00}

Early experimental works following this prediction did not give any
clearcut conclusion about the magnetic properties of (Ga,Mn)N:
ferromagnetic properties at high temperature,\cite{Hori02,Chitta04}
as well as paramagnetic properties down to low
temperature,\cite{Giraud04a,Wolos04a} have been reported. However,
there is now a general agreement that (Ga,Mn)N does not offer the
configuration which exists in (Ga,Mn)As, with the transfer of an
electron from the valence band to the $d$-levels, resulting in both
the $d^{5}$ configuration (with a spin 5/2) which is observed and
the acceptor character of the Mn impurity.

Ab initio calculations of the band
structure\cite{Kronik02,Kulatov02,Sanyal03,Popovic04,Wierzbowska04,Titov04}
do not exclude the possibility of a ferromagnetic interaction, but
they all conclude that the $d$~bands of Mn are located rather high
in the bandgap of GaN, although its exact position cannot be given.

From an experimental point of view, diverging conclusions have been
drawn about the valence state of Mn in (Ga,Mn)N. In bulk (Ga,Mn)N
with a very low Mn content, the $d^{5}$ state was detected by
electron paramagnetic resonance;\cite{Wolos04a} in similar samples,
the $d^{4}$ valence state was deduced from magneto-optical
measurements upon co-doping with Mg.\cite{Wolos04a} The Mn content
in these samples was chosen to be very low, a few
$10^{18}$~$\text{cm}^{-3}$ at most, so that these two results can be
understood from the strong $n$-type character of the first samples
and the usual acceptor character of Mg in GaN. The same $d^{5}$
state was observed by x-ray absorption spectroscopy at the $L$-edge
of Mn near the surface of layers grown by plasma-assisted molecular
beam epitaxy on n-type, Sn-doped GaN templates,\cite{Hwang05} while
we have observed the $d^{4}$ state by x-ray absorption spectroscopy
at the $K$-edge of Mn in layers grown also by plasma-assisted
molecular beam epitaxy but on undoped GaN templates.\cite{Titov04,
Sarigianidou06} Various types of valence states, from $d^{5}$ to
$d^{3}$, were invoked to explain optical spectra of epilayers with a
larger Mn content.\cite{Wolos04b,Han04} These results were mainly
obtained on (Ga,Mn)N with the wurtzite structure : the case of cubic
(Ga,Mn)N layers appears to be peculiar, since $p$-type doping due to
the incorporation of Manganese was reported,\cite{Novikov04,
Edmonds05}.

Optical absorption spectra\cite{Korotkov02,Graf02,Graf03,Wolos04b}
show (\emph{i}) a structured absorption band in the near infra-red,
with a sharp zero-phonon line, which was attributed to the $d-d$
transitions of the Mn$^{3+}$ ion (the $d^{4}$ configuration),
(\emph{ii}) a broad absorption band attributed to the transitions
between the Mn levels and the bands of the semiconductor,
(\emph{iii}) absorption at the bandgap of the semiconductor.

The present work is devoted to a magneto-spectroscopic study of
(Ga,Mn)N samples with a low Mn content, so that quite sharp features
are observed, and to the evolution of the magnetic and
magneto-spectroscopic properties upon increasing the Mn content.
Main results are a determination of the parameters which enter the
effective Hamiltonian of the ground level of Mn in (Ga,Mn)N, which
results from a dynamical Jahn-Teller effect in the $d^{4}$
configuration of Mn, and the observation of spin-carrier coupling
through the giant Zeeman effect at the bandgap of (Ga,Mn)N.

\section{Samples and experimental set-up}\label{sec:level1}
Epilayers of wurzite (Ga,Mn)N were grown by nitrogen-plasma-assisted
molecular beam epitaxy (MBE), on a GaN template, previously grown by
metal-organic chemical vapor deposition (MOCVD) on the $c$ surface
of a sapphire substrate. A few additional samples were grown on AlN
templates, so that the substrate is transparent in the vicinity of
the bandgap of (Ga,Mn)N. Unfortunately, the growth had to be
initiated with a  GaN layer, which was kept as thin as possible. The
growth temperature was kept at $720^{\circ}\text{C}$, which is our
usual growth temperature for GaN. A special attention was paid to
the different growth regimes related to the Ga/N flux ratio and to
the manganese flux.\cite{Giraud04b} Samples were characterized
$in$~$situ$ using reflection high-energy electron diffraction
(RHEED). Post-growth characterization comprised $ex$~$situ$
secondary ion mass spectroscopy (SIMS) and x-ray diffraction (XRD).

Optical absorption in the near infrared was measured using a Fourier
Transform Infrared spectrometer (FTIR) with a spectral resolution of
0.5~meV, using a tungsten lamp as the light source and a silicon
photodetector. Magneto-optical spectroscopy in the near infrared and
in the vicinity of the band gap energy was measured using a CCD
camera attached to a grating spectrometer. The magnetic circular
dichroism was then calculated from the transmission spectra of
right- and left-circular polarized light. In some cases, it was also
measured directly using a 50~kHz photoelastic modulator and a
photomultiplier tube.

Cathodoluminescence was observed in a scanning electron microscope,
with the sample placed on a cold finger at 5~K, using an
accelerating voltage of 4~kV and a typical current of the order of
30~nA. The luminescence intensity was recorded on a GaAs
photomultiplier tube attached to a 460~mm spectrometer for imaging,
and a charge coupled device (CCD) camera for spectroscopy. Attempts
to measure the photoluminescence excited by a He-Cd laser (20~mW on
a 50~$\mu$m spot size) were unsuccessful.

All the transmission spectra exhibited interference fringes due to
internal reflexions between the surface and the sapphire/GaN
interface. These interferences were removed using a lineshape fit of
the transmittance which includes a $\sin(2ne/\lambda)$ function,
where $e$ is the thickness of the GaN + (Ga,Mn)N layer, $\lambda$
the wavelength, and $n$ the optical index of GaN at room
temperature.\cite{Yu97} This process is quite efficient in the case
of sharp lines, such as the near infrared zero-phonon line, for
which the fit can be optimized over a narrow range across the line.
The process is less accurate for broad absorption bands (broader
than 25~meV typically), particularly in the vicinity of the bandgap,
since we should use the optical index of (Ga,Mn)N at low
temperature.

Magnetization measurements were performed using a superconducting
quantum interference device (SQUID) magnetometer at magnetic fields
up to 5~T applied in the plane of the sample ($\bot~c$~axis) or
perpendicular to the plane ($\|~c$ axis), at temperatures down to
2~K. The large diamagnetic contribution from the substrate was
evaluated on a piece of the same substrate.

\section{Magneto-optical spectroscopy of $d-d$ transitions}\label{sec:level1}
In this section, we present the spectroscopic study in the near
infrared, including magneto-transmission and cathodoluminescence.
Previous transmission spectra have been attributed\cite{Wolos04b} to
the $^{5}T_{2}\rightarrow \,^{5}E$ transition of the $d^4$
configuration of Mn, involving a large cubic crystal field
splitting, about 1.4~eV. The better resolution achieved on the
present samples with a low Mn content, and the comparison between
transmission and cathodoluminescence spectra, lead us to reconsider
the effective Hamiltonian which takes into account the Jahn-Teller
effect and the trigonal crystal field.

\subsection{Experimental results}\label{sec:level2}

\begin{figure}
\begin{center}
\includegraphics[scale=0.75]{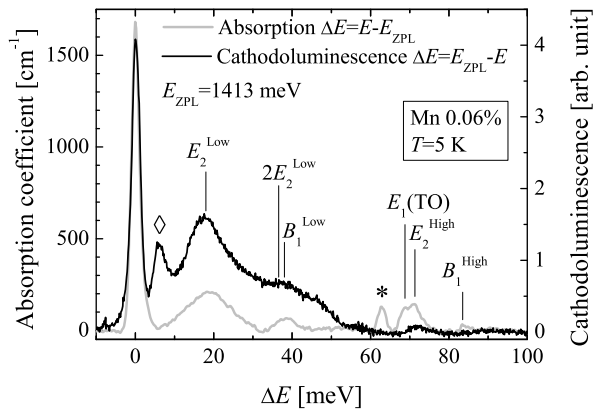}
\caption{Absorption and cathodoluminescence spectra of a (Ga,Mn)N
layer with 0.06\%~Mn. The horizontal scale is the distance from the
zero phonon line at 1413~meV. The position of high-symmetry phonons
of GaN\cite{Gebicki00, Harima04, Hasuike04} is indicated. Local
modes around the Mn impurity are expected at 20 and 70~meV. Diamond
and star indicate additional zero-phonon lines.} \label{fig1}
\end{center}
\end{figure}

Fig.~\ref{fig1} displays a transmission spectrum and a
cathodoluminescence spectrum of a (Ga,Mn)N layer with 0.06\%~Mn. A
series of cathodoluminescence spectra were measured locally from
different points of the sample. The one displayed in Fig.~\ref{fig1}
is typical of what was obtained from most points ; occasionally,
broadened spectra were observed. The origin of the horizontal scale
in Fig.~\ref{fig1} is the common position of the zero-phonon line at
1413~meV, and the scale is inverted for the two spectra.

Most features are present on both spectra : they are clearly phonon
replica. They may involve local modes of the Mn impurity and its
cluster of nearest neighbors, which are expected around 20~meV and
70~meV.\cite{Korotkov02} These frequencies are close to high
symmetry phonons of GaN,\cite{Gebicki00, Harima04, Hasuike04} as
expected since the difference of mass between Mn and Ga is small.
The presence of resonances with the phonons of GaN makes the
spectrum quite complex, but the coupling to the mode at 20~meV
dominates, so that a simple model of coupling to a single local mode
can be a reasonable approximation.

The coupling to this principal mode is rather weak: the intensity of
the zero-phonon line and that of the phonon replica are of the same
order of magnitude. In the case of a single local mode, the
intensity of the $n^{th}$ replica to the zero-phonon line is
expected to be $S^{n}/n!$, where $S$ is the so-called Huang-Rhys
factor: from the first replica, we obtain $S=$0.7 from the
transmission spectrum, $S=$1.9 from the cathodoluminescence
spectrum. This is in good agreement with previous determinations
($S=$0.6 in Ref.~\onlinecite{Wolos04b}, $S=$1.1 in
Ref.~\onlinecite{Korotkov02}). The main contribution to the line at
$\Delta$\textit{E}=40~meV arises from the second phonon line due to
the same mode.

This coupling is much weaker than in the case of Cr in II-VI
semiconductors,\cite{Vallin70} which has the same $d^{4}$
configuration. Then the zero-phonon line is weak ($3.7\times
10^{-27}~$m$^{2}~$eV/Cr in ZnSe,\cite{Vallin70} to be compared to
$1.1\times 10^{-23}~$m$^{2}~$eV/Mn in GaN). The intensity is
transferred to a broad vibronic band, three orders of magnitude more
intense than the zero-phonon line (so that the total integrated
intensity is about the same in both cases). The Huang-Rhys factor,
as deduced from the position of the maximum of the vibronic band, at
$S\hbar\omega$ from the zero-phonon line, is as high as 13 for
Cr$^{2+}$ in ZnTe and 18 in CdS.

Two lines in Fig.~\ref{fig1} cannot be ascribed to phonon replica.
The line marked with a diamond in Fig.~\ref{fig1} is present only in
the cathodoluminescence spectrum. It must correspond to a transition
to an excited state of the ground $^{5}T_{2}$ multiplet. Another
line, marked with a star in Fig.~\ref{fig1}, is observed in
transmission but not in cathodoluminescence. Its intensity is much
lower than that of the zero-phonon line at 1413~meV, and we
tentatively ascribe it to a spin-forbidden transition to one of the
$^{3}T_{1}$ or $^{1}A_{1}$ levels, which are indeed quite close for
such a value of the cubic crystal field.\cite{Tanabe}

\begin{figure*}
\begin{center}
\includegraphics[width=17cm]{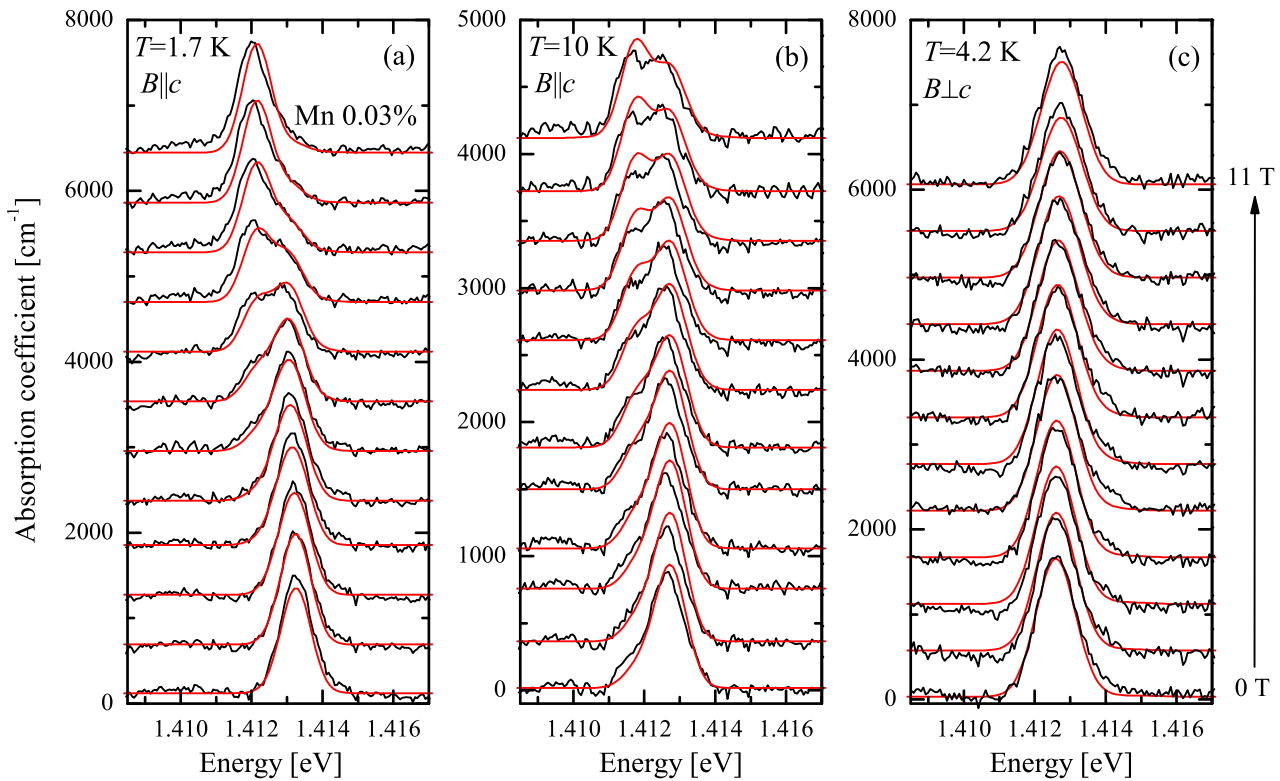}
\caption{(color online) Transmission spectra of a (Ga,Mn)N layer
with 0.03\%~Mn, for different fields applied in the Faraday
configuration at 1.7~K (a) and 10~K (b), and in the Voigt
configuration at 4.2~K (c). Black lines show the experimental
spectra and red lines the calculated ones.} \label{fig2}
\end{center}
\end{figure*}

Fig.~\ref{fig2} shows a series of transmission spectra in the range
of the zero-phonon line, for another sample with a low Mn content
(0.03\%), for different values of field and temperature, in both the
Faraday and Voigt configurations. The general features reported by
Wo{\l}os \textit{et al.}\cite{Wolos04b} are observed, with however a
higher resolution. Spectra up to 22~T have been also recorded from
the same sample, with a slightly lower resolution (not shown).

\begin{figure}
\begin{center}
\includegraphics[width=8.5cm]{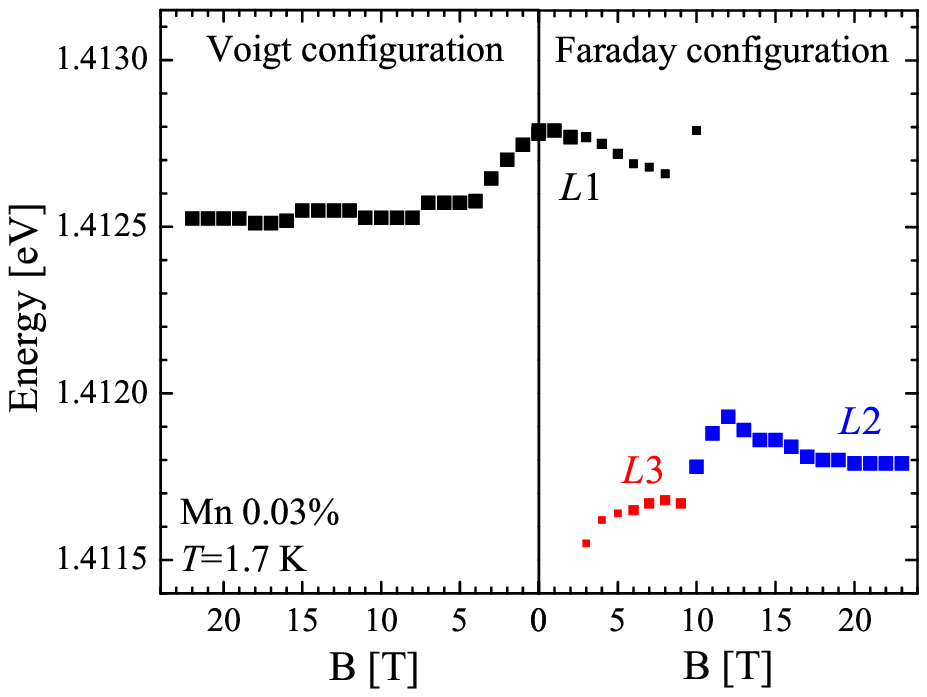}
\caption{(color online) Position of the absorption lines observed in
the (Ga,Mn)N layer with 0.03\%~Mn, as a function of the magnetic
field applied in the Faraday and the Voigt configurations. The size
of the symbols is proportional to the intensity. The sample is the
same as in Fig.~\ref{fig2} but the experiments were realized in a
different set-up.} \label{fig3}
\end{center}
\end{figure}

Two points should be stressed. First, the zero-field spectrum
contains essentially one line : any structure, if present, should be
comprised within the observed linewidth of 1.1~meV. Second, the
"shift" at 7~T reported in Ref.~\onlinecite{Wolos04b} is actually an
intensity transfer between two lines labeled \textit{L}1 and
\textit{L}2. From a plot of the positions and intensities deduced
from a Gaussian fit (Fig.~\ref{fig3}), a third peak $L3$ appears at
intermediate magnetic field, which exhibits a small but systematic
shift with respect to $L2$.

\begin{figure}
\begin{center}
\includegraphics[width=8.7cm]{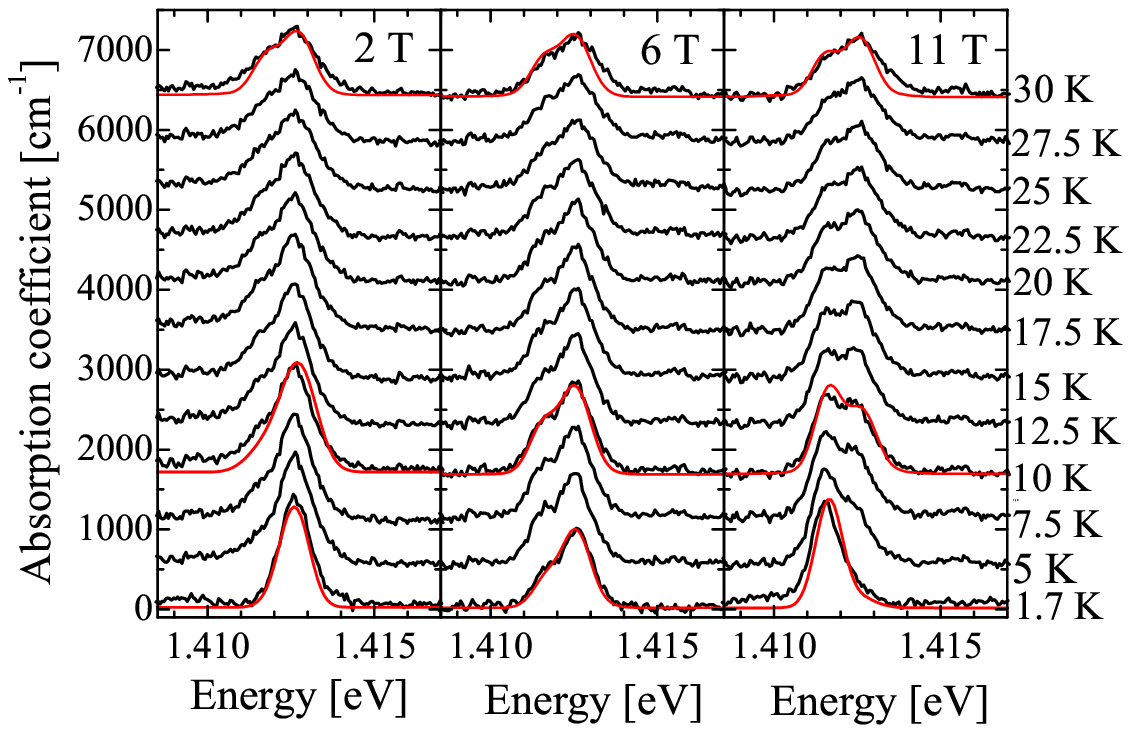}
\caption{(color online) Transmission spectra of the (Ga,Mn)N layer
with 0.03\%~Mn, for different fields applied in the Faraday
configuration, and different temperatures, as indicated. Black lines
show the experimental spectra and red lines the calculated ones.}
\label{fig4}
\end{center}
\end{figure}

The appearance of $L3$ at 3~T, on one hand, and the intensity
transfer to $L2$ observed between 10 and 13~T, on the other hand,
indicate two values of the applied field where a crossing takes
place in the manifold of ground states. The relative position of the
ground states which serve as initial states for the absorption lines
can be deduced from spectra recorded as a function of the
temperature for selected values of the magnetic field
(Fig.~\ref{fig4}). For instance, it is clear from the spectra that
two lines, at 1411.8~meV and at 1412.8~meV, co-exist even at low
temperature at 6~T. At lower fields only one line is seen at 1.7~K
and a second line rises on its low energy side at higher temperature
; the opposite is observed at higher fields. A plot of the
intensities (not shown) is well reproduced by an activation energy
with Land\'{e} factors of the order of 2. As the positions of $L1$,
$L2$ and $L3$ remain almost unchanged (at most 0.2~meV over a 10~T
scan), almost identical Zeeman shifts occur in both the ground and
excited multiplets - which may be accidental, but rather suggests
that the spin-orbit coupling is weak.

\begin{figure}
\begin{center}
\includegraphics[width=8cm]{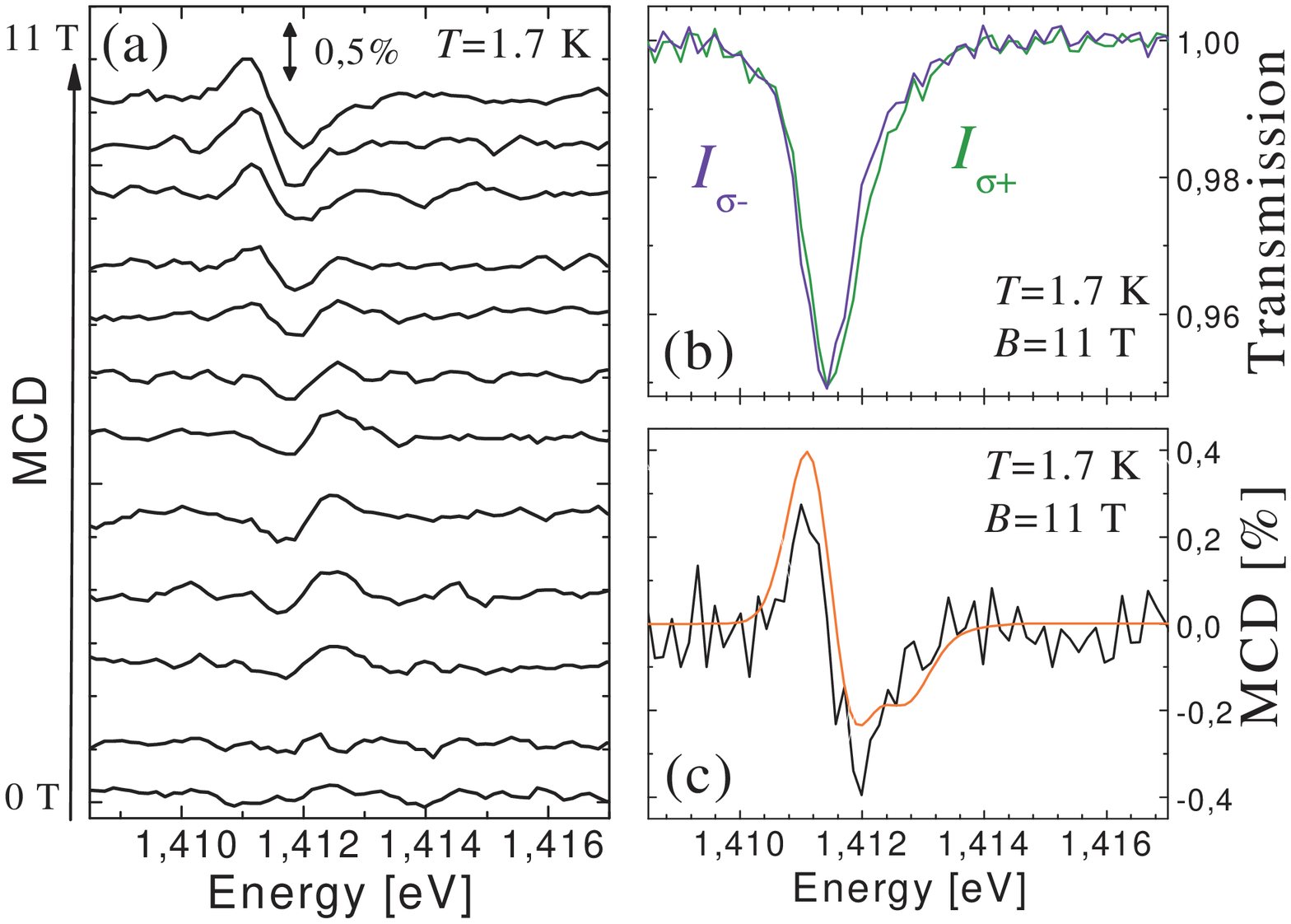}
\caption{(color online) (a) magnetic circular dichroism for
different values of the applied field (field step 1T)(b)
Transmission of right- and left-polarized light (green and blue
lines respectively) and (c) magnetic circular dichroism (black line)
of the (Ga,Mn)N layer with 0.03\%~Mn, at 1.7~K at 11~T in the
Faraday configuration. Red line shows the calculated MCD spectrum.}
\label{fig5}
\end{center}
\end{figure}

This is supported by the fact that we measure a weak (although non
zero) magnetic circular dichroism (MCD),
$[I(\sigma^{+})-I(\sigma^{-})]/[I(\sigma^{+})+I(\sigma^{-})]$.
Fig.~\ref{fig5}b displays spectra recorded for the two circular
polarizations, $\sigma^{+}$ and $\sigma^{-}$, at 11~T and 1.7~K,
together with the resulting MCD (Fig.~\ref{fig5}c). Measuring such a
small dichroism on a thin layer is difficult. MCD spectra with a
much higher signal-to-noise ratio, obtained on bulk material, are
given in Ref.~\onlinecite{Wolos04b}. The same lineshape is observed
at high field, but the MCD persists at lower field in
Ref.~\onlinecite{Wolos04b} while we observe a faster decrease and
even a trend to a change of sign (Fig.~\ref{fig5}a).

We shall show now, by constructing an effective Hamiltonian, that
these spectra are well explained by a \textit{d}$^4$ configuration,
in a crystal field of trigonal symmetry, with a dynamical
Jahn-Teller effect which reduces the spin-orbit coupling.

\subsection{Effective Hamiltonian}\label{sec:level2}

The Hamiltonian of a $d^{4}$ ion contains the crystal field terms
(cubic and trigonal components), the spin-orbit coupling, the Zeeman
effect, the coupling to local strain, and the Jahn-Teller effect.

The cubic component of the crystal field splits the $^{5}D$
multiplet into a ground $^{5}T_{2}$ orbital triplet and a $^{5}E$
orbital doublet (noted $10Dq$ in table I). A basis of orbital states
of the triplet adapted to the cubic symmetry is given by the three
states $|\xi>$, $|\eta>$, $|\varsigma>$ which transform like $yz$,
$zx$, $xy$ where $x$, $y$, $z$ label the cubic axes; for the
doublet, a basis is given by $|\theta>$, $|\varepsilon>$.

In the case of a wurtzite crystal, there is also a trigonal crystal
field. It has off-diagonal matrix elements in the cubic basis of the
orbital triplet and no matrix elements within the doublet. In the
case of epitaxial layers grown along the $c$ axis, it may contain a
contribution from the biaxial strain due to the partially
compensated lattice mismatch.

For the free Mn$^{3+}$ ion, the spin-orbit coupling within the
$^5$\textit{D} multiplet is written $\lambda \textbf{L}.\textbf{S}$
with $L=2$, $S=2$ and $\lambda\approx$~11~meV .\cite{Abragam} In a
crystal, this value can be  reduced slightly by hybridization, and
significantly by the Jahn-Teller effect (see below).

The coupling to local strain ("strain fluctuations") involves
trigonal and tetragonal components. The coupling to trigonal strain,
\textit{i.e.}, strain components of $T_2$ symmetry, has off-diagonal
matrix elements in the cubic basis of the orbital triplet and no
matrix elements within the doublet. The coupling to tetragonal
strain, \textit{i.e.}, strain components of $E$ symmetry, has matrix
elements within the triplet (where they are diagonal) and the
doublet.

Finally, there is a Jahn-Teller coupling of the orbital triplet to
$E$ and $T_{2}$ modes, and of the orbital doublet to the $E$ modes.

The most simple case of Jahn-Teller coupling is that of an orbital
triplet coupled to a single $E$ mode. A very intuitive treatment,
based on first and second order perturbations, was given by
Ham\cite{Ham65}. Basically, the orbital triplet is replaced by a
vibronic triplet : each orbital state is associated to a potential
well with its minimum corresponding to a distortion of the
environment along a cubic direction. The vibronic states are the
product of the orbital state by the eigenstates of the displaced
harmonic oscillator. The Hamiltonian acting within the orbital
triplet, as defined above, is replaced by an effective Hamiltonian
acting within the vibronic triplet: the main result is that the
first-order contribution of off-diagonal operators is reduced by the
overlap of eigenstates of the different harmonic oscillators.
First-order contributions of diagonal operators are not reduced.
Second-order contributions have to be considered : they are not
reduced as the first-order ones so that their role can be enhanced.

In the case of a crystal of cubic symmetry, a strong Jahn-Teller
effect results in a strong reduction of all contributions but that
of tetragonal strain fluctuations. One can use the approximation of
the "static Jahn-Teller coupling": the local tetragonal strain
slightly lowers the energy of a well corresponding to a distortion
along one of the three cubic directions, and for this ion all
calculations are performed with this particular distortion. This
approximation was used in the first study of Cr$^{2+}$ in II-VI
semiconductors,\cite{Vallin70} with the additional assumption that
there is no Jahn-Teller coupling in the $^{5}E$ excited state. The
same assumption was used later\cite{Herbich98} in the case of
Cr$^{2+}$ in CdS, which has the wurtzite structure. More recently
\cite{Wolos04b} a reasonable agreement with the available data was
obtained in the case of Mn$^{3+}$ in GaN.

In the present study, two points however cannot be explained by this
model: (\emph{i}) the presence of a second zero-phonon line in the
cathodoluminescence spectrum, and (\emph{ii}) the observation of a
main zero-phonon line at zero field, both in cathodoluminescence and
in transmission, with a linewidth of only 1~meV. Spectra calculated
within the static limit of the Jahn-Teller effect (see Fig.~8 of
Ref.~\onlinecite{Wolos04b}) show a doublet which could fit the
available spectra with a broad linewidth, but is ruled out by the
spectra of the present study (Fig.~\ref{fig1}). We could not obtain
a better fit with the same model and different parameters. This
leads us to re-examine the treatment of the Jahn-Teller effect in
the case of (Ga,Mn)N.

We may note that an improvement of the available spectra for
Cr$^{2+}$ in ZnS and ZnSe recently lead Bevilacqua \textit{et
al.}\cite{Bevilacqua04} to propose a rather weak Jahn-Teller
coupling to the $T_{2}$ modes (and not the $E$ mode) for the orbital
triplet, and a stronger coupling to the $E$ mode for the orbital
doublet (instead of no coupling). Both cases are more difficult to
handle, and necessitate a more complete treatment of the vibronic
states\cite{Bevilacqua04} which is beyond the scope of the present
study.

Actually, we obtained a reasonable, although not perfect, agreement
in the frame of a \emph{dynamic} Jahn-Teller effect for both the
triplet and the doublet, and not a static one for Cr in II-VI's.
This is not unexpected for two reasons: (\emph{i}) the smaller value
of the Huang-Rhys factor suggests that the Ham reduction
factor\cite{Ham65} should be smaller (this is but a hint, since the
Huang-Rhys factor measured on spectra involves a vibrational overlap
between a distorted harmonic oscillator of the triplet and a
distorted harmonic oscillator of the doublet, while the reduction
factors involve an overlap between two different harmonic
oscillators of two states of the same multiplet); (\emph{ii}) the
trigonal crystal field due to the wurtzite structure tends to mix
the three cubic states and, even if it is reduced by the Jahn-Teller
effect, it may remain larger than the tetragonal component of strain
fluctuations (which tends to stabilize one cubic distortion).

The effective hamiltonian which must be diagonalized within the
$^{5}T_{2}$ vibronic multiplet thus reads:
\begin{eqnarray}
H=\kappa V_{tri}
[3\widetilde{L}_{c}^{2}-\widetilde{L}(\widetilde{L}+1)]
-\kappa\lambda\widetilde{\textbf{L}}.\textbf{S}
+\mu_{B}\textbf{B}.(-\kappa
\widetilde{\textbf{L}}+2\textbf{S})\nonumber\\
+\rho(\widetilde{L}_{x}S_{x}\widetilde{L}_{y}S_{y}+perm.)+\textit{d}(\widetilde{L}_{x}^{2}S_{x}^{2}+perm.) \nonumber\\
\end{eqnarray}
where $V_{tri}$ measures the trigonal component of the crystal
field, $\widetilde{L}_{c}$ is the projection of the pseudo-kinetic
moment operator $\widetilde{L}=1$ on the $c$ axis, $\kappa$ is the
Ham reduction factor which applies on all first-order contributions
which are off-diagonal in the cubic basis, and the two last terms
arise from second-order contributions of spin-orbit coupling. We
assume that strain fluctuations have a negligible effect.

The case of the Jahn-Teller effect in the doublet is more
complicated. The full treatment for Cr$^{2+}$ in ZnS and
ZnSe\cite{Bevilacqua04} concludes that there is a significant
Jahn-Teller effect, and that the excited levels which contribute to
the absorption lines are comprised within a narrow energy range
(less than 0.3~meV). We keep a simple description and assume a
vibronic doublet, with an effective Hamiltonian parameterized by
$d'$ which contains a second-order contribution of spin-orbit
coupling (arising from the $\lambda_{TE}\textbf{L}.\textbf{S}$
spin-orbit coupling between the ground triplet and the orbital
doublet), and the Zeeman effect. Using the operators
$\textbf{U}_{\theta}$ and $\textbf{U}_{\varepsilon}$ defined by
Ham\cite{Ham69} within the orbital doublet, we have
\begin{eqnarray}
H=d'[U_{\theta}(2S_{z}^{2}-S_{x}^{2}-S_{y}^{2})+U_{\varepsilon}.\sqrt{3}(S_{x}^{2}-S_{y}^{2})]
\nonumber\\+\mu_{B}\textbf{B}.2\textbf{S}
\end{eqnarray}

\begin{figure}
\begin{center}
\includegraphics[width=8.7cm]{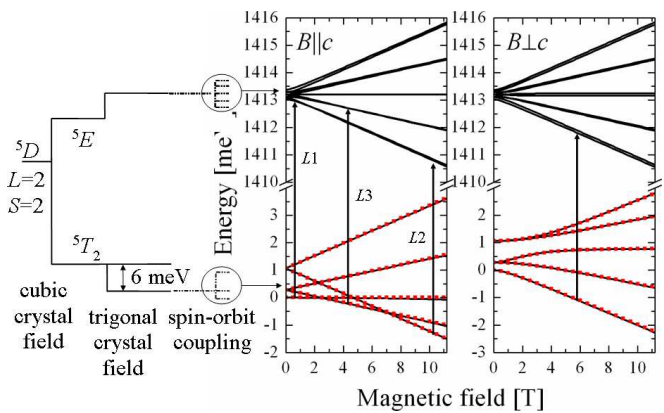}
\caption{Energy levels calculated taking into account the cubic and
trigonal crystal field, the spin-orbit coupling, the Zeeman effect
in both Faraday and Voigt configuration and the Jahn-Teller effect.
The arrows indicate the main components of the optical transitions.
The red dotted lines correspond to a fit of the energy levels of the
ground state using the effective spin Hamiltonian (4). The
corresponding fit parameters are given in table II} \label{fig6}
\end{center}
\end{figure}

The intensity of the $^{5}T_{2}\rightarrow \,^{5}E$ transitions is
proportional to the square of the matrix element of the electric
dipole, which in a tetrahedral configuration is itself proportional
to the matrix element of the appropriate projection of \textbf{L}.
Spectra were calculated by adding all contributions:
\begin{equation}
I_\pm=\frac{\sum_{i,f}I_{0}~\exp(-(E-E_{if})^{2}/2\sigma^{2})~|\langle i|L^{\pm}|f\rangle|^{2}~n(E_{i})}{\sum_i n(E_i)}
\end{equation}

For absorption, $|i>$ spans all calculated states of the $^{5}T_{2}$
multiplet, taking into account the thermal occupancy
\textit{n}(\textit{E}$_i$)=e$^{-E_i/k_BT}$, and $|f>$ all calculated
states of the $^5$\textit{E} multiplet. The full width at half
maximum of the Gaussian lineshape (equal to 0.8~meV) and its
intensity were adjusted at 1.7~K and zero-field, and kept constant
for all temperature and field values.

Absorption and emission spectra were calculated using the parameters
listed in table \ref{Parameters}. Fig.~\ref{fig6} shows the energy
levels. The position of the second zero phonon line observed in
cathodoluminescence 6~meV above the main line corresponds to the
splitting of the orbital triplet by the trigonal crystal field.
Calculated transmission spectra are shown by red lines in
Fig.~\ref{fig2}, \ref{fig4} and \ref{fig5}. All lines contain
several components: the main contributions are shown by arrows in
Fig.~\ref{fig6}. The observed "jumps" are due to the crossing
between the spin states of the ground orbital singlet. The
$g$-factors in the ground and excited multiplets are close to 2 so
that the Zeeman shifts are parallel. The position in field of the
crossings is directly related to the zero-field splitting, due to
the combined effect of the trigonal crystal field and the spin orbit
coupling. From Fig.~\ref{fig2} and \ref{fig4}, the model of Mn in
the \textit{d}$^4$ configuration well describes the field and
temperature dependence of the absorption spectra. The MCD at high
magnetic field is also well described (Fig.~\ref{fig5}).

The calculated spectra deviate from our experimental findings on two
points. First, the intensity of the second zero-phonon line in
emission (labeled by a diamond on Fig.~\ref{fig1}) is about twice
larger that experimentally observed in cathodoluminescence. Second,
the calculated MCD is not dramatically reduced when the field is
decreased : this is in good agreement if compared to the observation
of Wo{\l}os et al. \cite{Wolos04b}, but our observations (although
with a low signal-to-noise ratio) tend to indicate a decrease and a
change in shape of the MCD at small field. Note that the MCD is
small indeed, and as such, may be sensitive to fluctuations of the
trigonal crystal field, which can be different in a bulk crystal and
an epitaxial layer.

\begin{table}[ht]
\center{
\begin{tabular}{c c|c}
  Cubic cristal field & 10\textit{Dq} & 1407~meV\\
  \hline
  \hline
  Trigonal cristal field & $3\kappa$\textit{V}$_{tri}$ & 6.4~meV\\
  \hline
  \hline
  Spin-orbit & $\lambda_{eff}=\kappa\lambda$ & 1.1 meV\\
  coupling & $\rho$ & 0.6 meV\\
  of $^5T_2$ state & $|d|$ & 0.02 meV\\
  \hline
  of $^5E$ state & $|d'|$ & 0.015 meV\\
  \hline
  \hline
  Ham reduction factor & $\kappa$ & 0.1\\
\end{tabular}
\caption{\label{Parameters}Parameters of the effective Hamiltonian.}}
\end{table}

In the perturbation treatment of the dynamic Jahn-Teller effect, the
ratio $\lambda_{eff}$/$\lambda$ is equal to the Ham reduction
factor: hence $\kappa\approx0.1$. As $\kappa$=e$^{3\textit{S}/2}$,
the Huang-Rhys factor is \textit{S}=1.5, in reasonable agreement
with the values describing the intensity of the one-phonon replica
in absorption and cathodoluminescence (0.7 and 1.9 respectively).

The other parameters, $\rho$ and $d$, govern two components of the
effective Hamiltonian which are built through symmetry arguments.
They can be estimated as second-order terms in the perturbation
scheme. For instance, we expect a second-order contribution of the
spin-orbit coupling through the excited vibrational states of the
ground triplet,\cite{Ham65}, another contribution through the $E$
doublet (the $\lambda_{TE}$ term of Ref.~\onlinecite{Herbich98}),
which is opposite in sign, and a third contribution from spin-orbit
coupling through excited triplet states and spin-spin
coupling.\cite{Vall74} Note however that this approach (perturbation
treatment with coupling to a single $E$ mode) is oversimplified and
a full calculation, possibly  involving also a coupling to
\textit{T}$_2$ mode \cite{Bevilacqua04}, would be necessary. This
calculation, performed for (Zn,Cr)S and (Zn,Cr)Se, has even shown a
stronger coupling to \textit{T}$_2$ mode than \textit{E} mode.

Nevertheless, the description of the magneto-optical data has been
improved by introducing a simple description of the dynamic
Jahn-Teller effect.

It may be also useful to define a spin Hamiltonian acting within the
ground orbital singlet, which is separated by 6~meV from the doublet
by the trigonal field. Its form for a spin 2 can be found in
Ref.~\onlinecite{Graf03, Vall74, Krei96}. In a wurtzite
semiconductor, it remains : \\
\begin{eqnarray}
H=(g_{\|}-g_{\bot})\mu_B B_cS_c+g_{\bot}\mu_B\textbf{B}.\textbf{S}\nonumber\\
+D[S_c^2-\frac{1}{3}S(S+1)]\nonumber\\
+\frac{1}{180}F\{35S_c^4+[25-30S(S+1)]S_c^2+72\}\nonumber\\
+\frac{1}{6}a(S_x^4+S_y^4+S_z^4-\frac{102}{5})
\end{eqnarray}
As above, $c$ labels the $[000.1]$ axis of the wurtzite structures,
and $(x,y,z)$ the cubic axes associated with the tetrahedron of
nearest neighbors.

The energy levels of the ground state calculated with the effective
Hamiltonian (1) can be well reproduced using the spin Hamiltonian
(4), as shown by the red dotted lines in Fig.~\ref{fig6} calculated
for the parameters given in table II.
\begin{table}[ht] \center{
\begin{tabular}{c|c|c}
  & Mn$^{3+}$ (3$d^4$) S=2 & Mn$^{2+}$ (3$d^5$) S=5/2 \\
  \hline
  \hline
   $g_{\|}$ & 1.91 & 1.9994 \\
   $g_{\bot}$ &1.98& 1.9994 \\
  $D$& 0.27 meV & -2.5 $\mu$eV \\
  $a-F$& $\approx$ 0.03 meV & $\approx$ 0.07 $\mu$eV \\
\end{tabular}
\caption{Spin Hamiltonian parameters describing the ground spin
multiplet of a Mn$^{3+}$ (3d$^4$) ion in GaN. The corresponding
parameters for a Mn$^{2+}$ (3d$^5$) in GaN\cite{Graf03} are given
for comparison.}}
\end{table}
We obtain $g$-factors slightly smaller than 2 for both
crystallographic directions (the orbital contribution is reduced by
the Jahn-Teller effect), and a large axial anisotropy parameterized
by $D$, which dominates the fine structure. As for Mn$^{2+}$,
\cite{Graf03} the cubic anisotropy terms $a$ and $F$ are much
smaller than the axial term $D$. From the energy levels calculated
in zero field with (1), a rough estimation of $a-F\approx$ 0.03 meV
can be obtained. The fine structure parameters have the same order
of magnitude than the ones measured for Cr$^{2+}$(3$d^4$) ions in
II-VI's. \cite{Vall74} It should be noted that these parameters are
much larger than the corresponding ones determined by EPR for
Mn$^{2+}$ ions in GaN (3$d^5$ configuration).\cite{Graf03}

We can also calculate the magnetization of isolated Mn ions. The
result is not much different from the average magnetization
calculated under the assumption of a static Jahn-Teller effect.
\cite{Gosk} As expected, it is strongly anisotropic, the
\emph{c}-axis being a hard axis.

\begin{figure}
\begin{center}
\includegraphics[width=8.7cm]{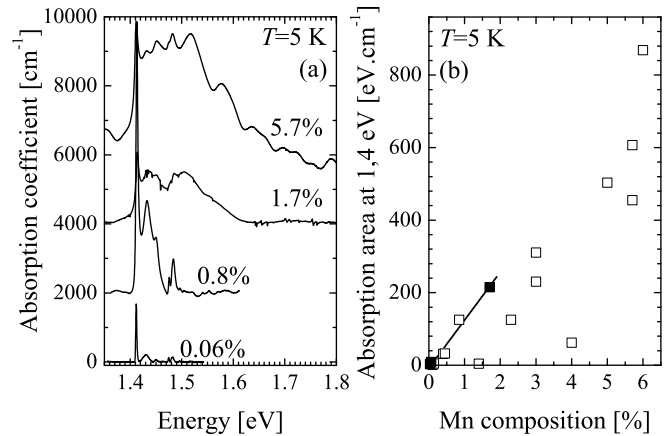}
\caption{(a) Absorption spectra for 0.06, 0.8, 1.7 and 5.7\% Mn in
GaN layers (b) area of the absorption band integrated over the
energy range shown in (a), as a function of the Mn content. The
samples with 0.03, 0.06 and 1.7 \% Mn which have been studied in
more details are shown by closed symbols.} \label{fig7}
\end{center}
\end{figure}

The very sharp spectra described above have been obtained on samples
with a very low Mn content, typically less than 0.1\%. Very similar
spectra, although broadened, have been observed up to 1\%~Mn (see an
example in Fig.~\ref{fig7}a). The intensity of the whole structure
is proportional to the Mn content determined by SIMS (straight line
in Fig.~\ref{fig7}b). We could follow the absorption band in the
near infrared up to 6\%~Mn, using FTIR in zero field. The strong
broadening considerably smooths the structure, and finally it
becomes difficult to eliminate the effect of the interferences and
to determine the lineshape with a reasonable accuracy (top of
Fig.~\ref{fig7}a). Nevertheless, the integrated area of the
absorption band keeps increasing with the Mn content (Fig.~
\ref{fig7}b), showing that in most samples the majority of the Mn
ions is in the \textit{d}$^4$ configuration.

\section{Magneto-optical spectroscopy at the bandgap}\label{sec:level1}
Magnetic circular dichroism at the band gap edge has been studied
with the magnetic field applied parallel to the \textit{c}-axis. For
these studies, 0.5~$\mu$m thick (Ga,Mn)N epilayers have been grown
on AlN MOCVD buffer layers deposited on sapphire. A 100~nm thick GaN
buffer layer had to be grown first in order to adjust the growth
conditions. Here we describe the results obtained on a sample with
1.7\%~Mn.

\begin{figure}[ht]
\begin{center}
\includegraphics[width=8cm]{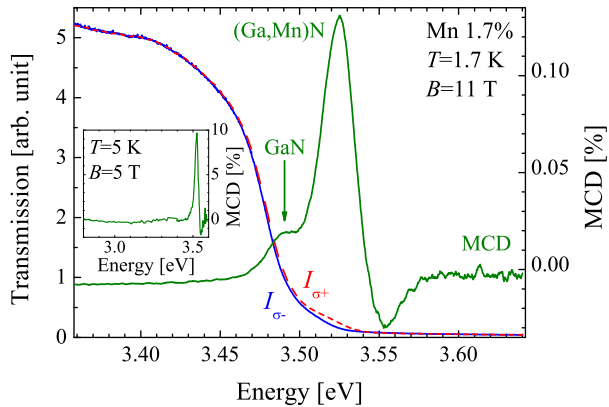}
\caption{(color online) Transmission of right- and left-polarized
light (red dash line and blue solid line respectively) of the
(Ga,Mn)N layer with 1.7\%~Mn, and magnetic circular dichroism
spectrum at 11~T in the Faraday configuration (green solid line).
The inset shows the MCD at 5~T with a broader energy scan.}
\label{fig8}
\end{center}
\end{figure}

Fig.~\ref{fig8} shows the transmission for $\sigma^{+}$ and
$\sigma^{-}$ polarized light, measured at 1.7~K and 11~T using a
linear polarizer and a Babinet compensator set as a quarter
retarding plate.

A broad absorption which rises with the Mn content is observed above
2.1~eV. \cite{Ferrand}. It was attributed to a band-level
transition.\cite{Wolos04a, Graf02} At higher energy, the band edge
rises in two steps: a weakly polarized one at low energy, which we
shall ascribe to the (albeit thin) GaN buffer layer; a second one,
with a clear circular polarization,\cite{Ferrand} which we shall
show to be due to the "giant Zeeman effect" in (Ga,Mn)N. Note the
unusual polarization which is observed, with the $\sigma^{-}$ edge
shifted to lower energy, in agreement with Ref.~\onlinecite{Ando}.

Fig.~\ref{fig8} shows also the MCD,
$[I(\sigma^{+})-I(\sigma^{-})]/[I(\sigma^{+})+I(\sigma^{-})]$. The
broad absorption band observed\cite{Ferrand} between 2.1 and 3.5~eV
shows no measurable polarization (inset of Fig.~\ref{fig8}): MCD is
only observed in resonance with the bandgap edge.

\begin{figure}[ht]
\begin{center}
\includegraphics[width=8cm]{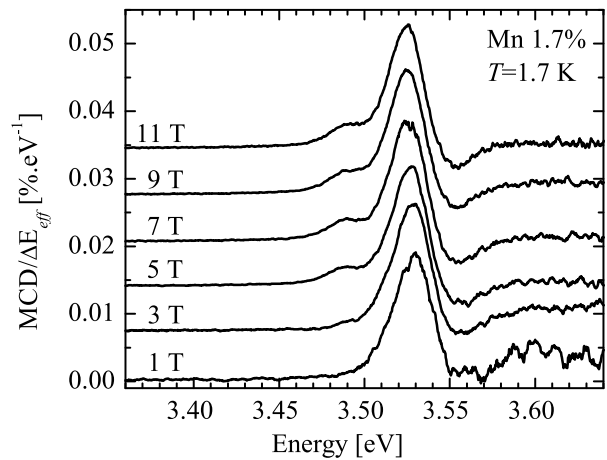}
\caption{Magnetic circular dichroism divided by the effective Zeeman
splitting $\Delta E_{eff}$, for the (Ga,Mn)N sample with 1.7\%~Mn at
1.7~K.} \label{fig9}
\end{center}
\end{figure}

\begin{figure}[ht]
\begin{center}
\includegraphics[width=8cm]{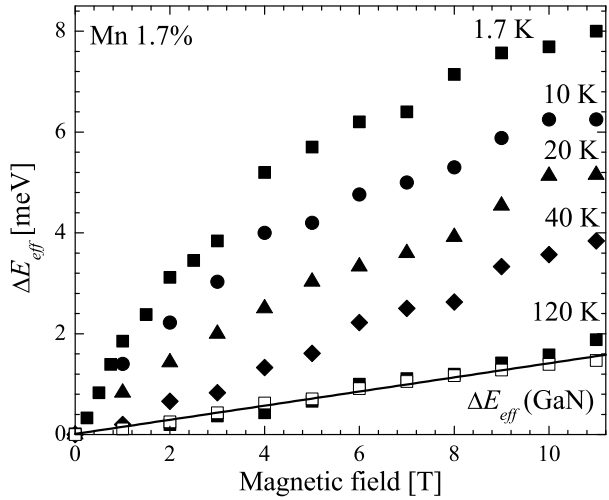}
\caption{Magnetic field dependence at 1.7~K of the effective Zeeman
splitting $\Delta E_{eff}$ observed at the bandgap of pure GaN (open
symbols) and of (Ga,Mn)N with 1.7\%~Mn (closed symbol).}
\label{fig10}
\end{center}
\end{figure}

If we assume that the absorption edge rigidly shifts in opposite
directions in $\sigma^{+}$ and $\sigma^{-}$ polarizations, the MCD
can be expressed as: \cite{Ando}
\begin{equation} \label{equation1}
\textrm{MCD}=-\frac{\textrm{dln}(T(E))}{\textrm{d}E}\frac{\Delta E}{2},
\end{equation}
with $T(E)$ the zero-field transmission measured with non-polarized
light at energy $E$, and $\Delta E=E(\sigma^{+})-E(\sigma^{-})$ the
Zeeman splitting. In the absence of any identified excitonic
structure, we shall use this expression to extract an
\emph{effective} Zeeman splitting $\Delta E_{eff}$. A minimal
requirement is that the ratio MCD$/\Delta E_{eff}$ should not depend
on the applied field. This is checked in Fig.~\ref{fig9} for the
high-energy part of the MCD (3.50 to 3.55 eV) observed at different
values of the applied field, which indeed exhibits a constant shape,
so that we can extract a value of $\Delta E_{eff}(H)$. This value is
plotted in Fig.~\ref{fig10}, together with the value measured on a
similar layer of pure GaN. In GaN, $\Delta E_{eff}(H)$ is linear in
field and does not depend on the temperature. In (Ga,Mn)N, it
increases non-linearly with the applied field, and decreases if the
temperature increases. Above 100~K, it assumes the same value as in
pure GaN. Note also in Fig.~\ref{fig9} that the low energy part of
the MCD spectrum, at 3.49~eV, keeps increasing at high field: this
part is actually characterized by a different effective Zeeman
splitting (not shown), equal to that of GaN. Hence we attribute this
signal to the thin GaN buffer layer.

If the effective Zeeman splitting we observe contains a contribution
due to the giant Zeeman effect, \emph{i.e.}, to the interaction
between the bands of the semiconductor and the localized spins of
Mn, it is expected to be related to the magnetization of the Mn
system, and to exhibit a specific dependence on the Mn content and
on the field and temperature. This we check now.

\begin{figure}
\begin{center}
\includegraphics[width=8.5cm]{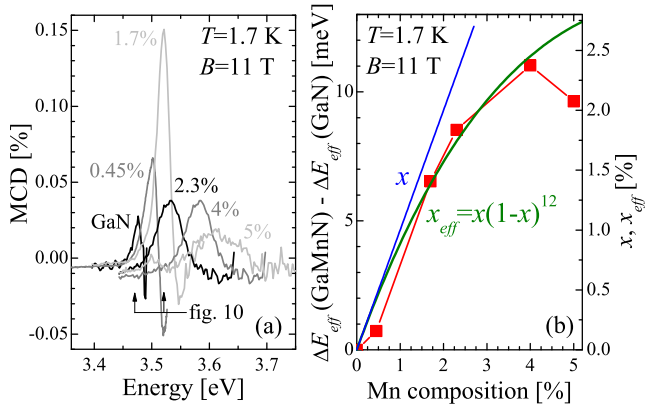}
\caption{(color online) (a) MCD spectra at 1.7~K and 11~T for
(Ga,Mn) layers with different values of the Mn content from 0 to
5\%, the arrows indicate the photon energies used in
Fig.~\ref{fig10} ; (b, symbols, left scale) $\Delta
E_{eff}(x)-\Delta E_{eff}(0)$ as a function of the Mn content
\textit{x} ;(right scale) density of Mn ions ($x$, blue solid line)
and of isolated Mn ions ($x_{eff}$, green line).} \label{fig11}
\end{center}
\end{figure}

Fig.~\ref{fig11}a shows MCD spectra at 1.7~K and 11~T for (Ga,Mn)N
layers with different values of the Mn content, from 0 to 5\%. The
MCD peak shifts to the blue upon increasing the Mn content. It shows
a strong resonance which markedly increases up to 2\%, then one
observes a broadening of the band edge and of the MCD signal, and a
decrease of the MCD peak intensity. However, the effective Zeeman
splitting $\Delta E_{eff}$ keeps increasing, as shown in
Fig.~\ref{fig11}b, which displays the change of $\Delta E_{eff}$ due
to the incorporation of Mn, measured at 1.7~K and 11~T. The
dependence on $x$ is sublinear, and actually quite close to the
variation of the density of free spins, $x_{eff}$, which would be
expected if nearest-neighbor Mn-Mn pairs are blocked by a strong
antiferromagnetic interaction. We shall come back to this point
later.

\begin{figure}
\begin{center}
\includegraphics[width=8cm]{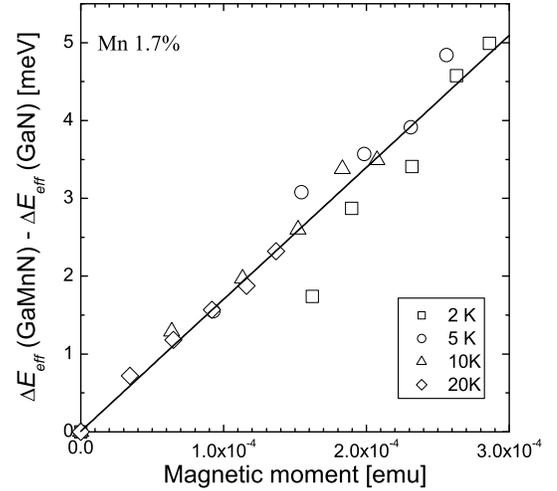}
\caption{Giant Zeeman splitting $\Delta E_{eff}(x,H,T)-\Delta
E_{eff}(x=0,H)$, plotted versus the magnetic moment measured by
SQUID between 0 and 5T with the field applied along the $c$-axis
(the line is a guide for eyes). The giant Zeeman splitting is found
to be proportional to the magnetization in the whole temperature and
magnetic field range.} \label{fig12}
\end{center}
\end{figure}

Fig.~\ref{fig10} suggests a paramagnetic-like behavior of the
Mn-induced change of $\Delta E_{eff}$ for the 1.7\% Mn layer. For
this sample, the giant Zeeman splitting can be determined by
subtracting the Zeeman splitting of the reference GaN layer.
Fig.~\ref{fig12} shows the giant Zeeman splitting, plotted
\emph{versus} the magnetic moment, measured by SQUID with the field
applied along the $c$-axis. In the whole temperature (2K-20K) and
field range (0-5T), the giant Zeeman splitting is found to be
proportional to the magnetization. This, and the dependence on the
Mn content, clearly evidence a coupling between the
\textit{s,p}-electrons of the host semiconductor and the
\textit{d}-electrons localized on the Mn$^{3+}$ ions.

Of course, a determination of the strength of the spin-carrier
coupling would be highly valuable. This is a rather simple question
in the case of Mn-based II-VI DMS's, involving  spins of high
symmetry (Mn$^{2+}$) in a zinc-blende semiconductor with moderate
excitonic effects. In the case of a $d^{4}$ magnetic impurity, the
$p-d$ coupling involves not only the usual spin-spin term, but also
an orbital coupling which has been particularly studied for
Cr$^{2+}$ in ZnSe.\cite{Bhattacharjee,Bhattacharjee1} We may expect
this term to be quenched by the Jahn-Teller factor. The influence of
the excitonic structure however cannot be neglected, as shown by the
analysis of the giant Zeeman effect measured in
(Zn,Co)O,\cite{Pacuski} another wide-gap DMS with the wurtzite
structure and a small spin-orbit coupling, or by the analysis of
stress effects in GaN epilayers.\cite{Gil} In both cases, the
electron-hole exchange interaction leads to anticrossings of the
various excitons ($A$, $B$ and $C$) of the wurtzite structure, so
that the shift of the exciton can be quite different from the shift
of the involved carriers.

Because of the broadening of the excitons, and of the overlap with
the contributions from the GaN buffer layer, excitonic transitions
are not observed in the present layers, neither in reflectivity nor
in transmission. Hence a quantitative analysis is out of reach.
Nevertheless, the MCD observed in transmission can be tentatively
attributed to the exciton of lowest energy, which in GaN is the
so-called \textit{A} exciton of the wurtzite structure.\cite{Gil}
The giant Zeeman splitting of the corresponding electron-hole pair,
in the Faraday configuration with the field along the $c$-axis,
is\cite{Pacuski}
\begin{equation}\Delta
E=E(\sigma^{+})-E(\sigma^{-})=N_0(\alpha-\beta)x<S_z>,
\end{equation}
where $\alpha$ and $\beta$ are the \textit{s}- and
\textit{p}-\textit{d} coupling constants, respectively, $N_{0}x$ is
the spin density, and $<S_c>$ is the spin component of Mn along the
$c$-axis (=-2 at saturation). From the spectra shown in Fig.
\ref{fig8}, a negative sign of $\alpha-\beta$ is inferred. This is
opposite to what was found in most DMS's studied so far, but in
Cr-doped II-VI semiconductors, \cite{Mac,Herbich98} another
realization of the $d^{4}$ configuration. As $\alpha$ is usually
positive, and small with respect to $\beta$, that means a
ferromagnetic interaction between the holes in the valence band and
the localized spins, as it has been predicted for DMS's with a less
than half filled \textit{d}-shell.\cite{Bhattacharjee1}

Finally, it should be stressed that, in the absence of a complete
identification of the various excitons, a quantitative analysis of
the giant Zeeman splitting would be adventurous. In particular, we
expect the excitonic shifts to show a complex dependence on the
magnetization, with anticrossings as soon as the exchange splitting
of the bands becomes of the order of the zero-field excitonic
splitting.\cite{Gil,Pacuski} That will cause the MCD to saturate,
which could contribute to the saturation observed in
Fig.\ref{fig11}b as a function of the Mn content, or in
Fig.\ref{fig10} as a function of the applied field (while the
magnetization calculated for a single Mn ion is still far from
saturation).\cite{Gosk}

\section{Conclusion}\label{sec:level1}
A detailed study, by magneto-optical spectroscopy in the near
infrared, of wurtzite (Ga,Mn)N layers grown by plasma-assisted
molecular beam epitaxy, confirms the substitution of Ga by Mn with
the $d^4$ configuration (Mn$^{3+}$). Very sharp spectra obtained on
samples with a low Mn content, and cathodoluminescence spectra, are
reasonably well understood using a crystal field model with a
dynamic (not static) Jahn-Teller effect.

Samples with a slightly larger Mn content show a MCD signal in
transmission, resonant with the bandgap edge. The effect increases
with the Mn content and behaves as the measured magnetization. This
evidences a coupling between the Mn$^{3+}$ \textit{d}-electrons and
the carriers of the host semiconductor ("giant Zeeman effect") ; the
sign of the MCD agrees with a ferromagnetic \textit{p-d}
interaction.

\begin{acknowledgments}
This work was supported by the E.~U. contract FENIKS
(G5RD~2001~00535). We thank Edith Bellet-Amalric for the structural
characterization of the samples.
\end{acknowledgments}

\end{document}